\input{aipcheck}

\documentclass[
    ,final            
  ]
  {aipproc}

\layoutstyle{6x9}

\begin{document}

\title{Properties of the intermediate type of gamma-ray bursts}


\author{I. Horv\'ath}{
  address={Department of Physics, Bolyai Military University, Budapest,
Box-12, H-1456, Hungary}
}

\author{F. Ryde }{
  address={Stockholm Observatory, AlbaNova, SE-106 91 Stockholm, Sweden}
}

\author{L.G. Bal\'azs }{
  address={Konkoly Observatory, Budapest, Box-67, H-1525, Hungary} 
}

\author{Z. Bagoly}{
  address={Laboratory for Information Technology, E\" otv\" os
          University, Budapest, P\'azm\'any P. s. 1/A,, H-1117,
          Hungary} 
}

\author{A. M\'esz\'aros }{
  address={Astronomical Institute of the Charles University,
              V Hole\v{s}ovi\v{c}k\'ach 2, CZ 180 00 Prague 8,
          Czech Republic} 
}

\begin{abstract}
 Gamma-ray bursts can be divided into three groups ("short",
"intermediate", "long") with respect to their durations.
The third type of gamma-ray bursts - as known - has
the intermediate duration.  We
show that the intermediate group is the softest one.
An anticorrelation between the hardness and the duration
is found for this subclass in contrast to the short
and long groups.

\end{abstract}

\maketitle


\section{Two-dimensional Gaussian fits}

  Simultaneously Mukherjee et al. \cite{muk98} and Horváth \cite{ho98} found a third 
group of gamma-ray bursts (GRBs). Somewhat later several authors \cite{hak,bala,rm, ho06} 
also suggested the existence 
of the third ("intermediate") group as well. The physical existence of the 
third group is, however, still not convincingly proven. For example, 
Hakkila et al. \cite{hak} believe that the third group is only a deviation 
caused by a complicated instrumental effect, which can reduce the durations 
of some faint long bursts. Later Hakkila et al.  \cite{hak03}
published another paper which had different conclusions. 

Using Principal Component Analysis (PCA), Bagoly et al. \cite{bag98} 
have shown that there are only two major quantities necessary 
(called the Principal Components; PCs) to characterize most of  
the properties of the bursts in the BATSE Catalog.  Consequently, 
the problem of the choice of the relevant parameters describing GRBs 
is basically a two-dimensional problem. For the statistical analysis 
the choice of two independent parameters is enough; they may be, but 
are not necessarily, the two principal components. This means that 
only two parameters, relevantly chosen, should be enough for the 
classification and determination of the groups. Here we have chosen {\bf duration} 
$T_{90}$ and {\bf hardness} $H_{32}=F_3/F_2$ ($F_3$ and $F_2$ are the fluences)
 for these parameters.

We can assume that the observed probability distribution of
GRBs in this plane is a superposition of the distributions
characterizing the different types of bursts present in the
sample. Introducing the notations $x= \log T_{90}$ and $y=\log
H_{32}$ and using the law of full probabilities we
can write
\begin{equation}\label{lfpr}
    p(x,y)
    =\sum \limits_{l=1}^k p(x,y|l)p_l.
\end{equation}


\begin{figure}
\includegraphics[height=.7\textheight, angle=270]{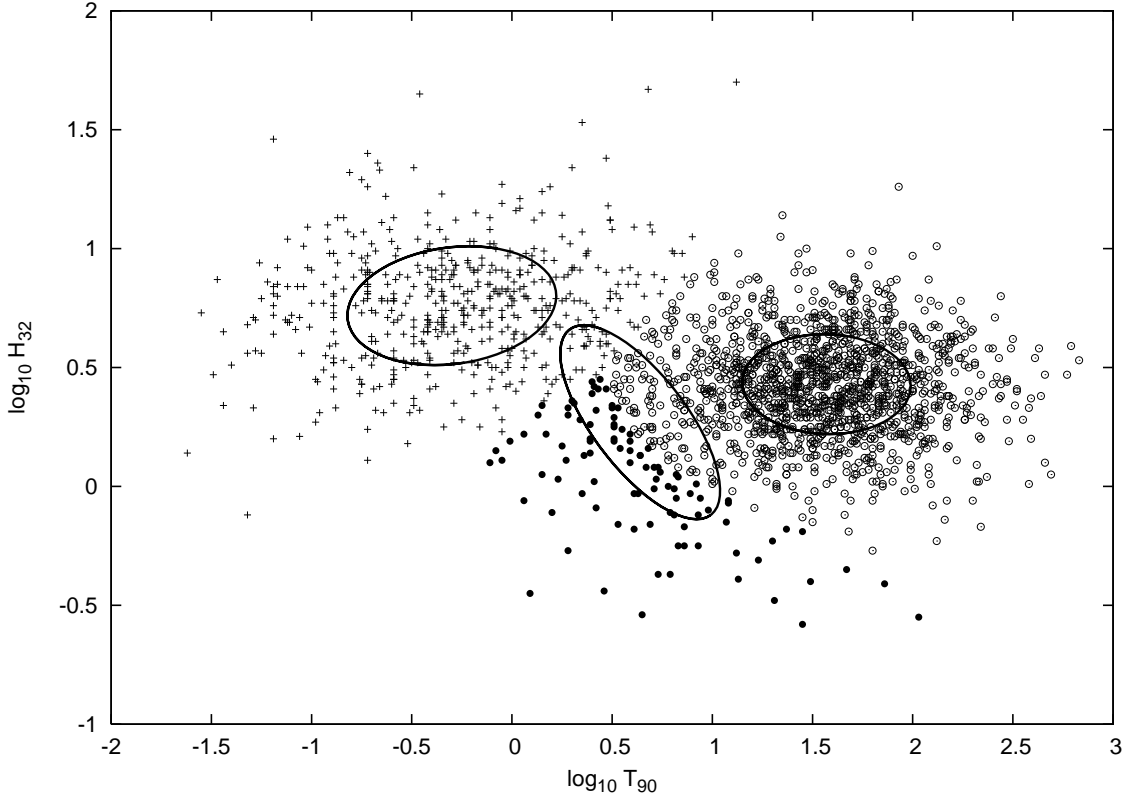}
\caption{The $ 1956$ GRBs in the $\{\log T_{90}; \log H_{32}\}$
plane. The $1\sigma$  ellipses of the three Gaussian distributions are
also shown, which were obtained in the ML procedure. The different
symbols (crosses, filled circles and open circles) mark bursts
belonging 
to the short, intermediate and long classes, respectively.}
\end{figure}

In this equation $p(x,y|l)$ is the conditional probability density assuming 
that a burst belongs to the $l$-th class. $p_l$ is the probability for this class 
in the observed sample, where $k$ is the number of classes. In order to 
decompose the observed probability distribution $p(x,y)$ into the 
superposition of different classes we need the functional form of $p(x,y|l)$. 
The probability distribution of the logarithm of durations can be well 
fitted by Gaussian distributions, if we restrict ourselves to the short 
and long GRBs, respectively \cite{ho02}. We assume the same for the 
$y$ coordinate as well. Therefore it holds

\begin{equation} \label{gauss}
p(x,y|l)  =
 \frac{(1-r^2)^{-{\frac{1}{2}}}}{2 \pi \sigma_x \sigma_y} 
\exp\left[-\frac{1}{2(1-r^2)}
\left(\frac{(x-a_x)^2}{\sigma_x^2} + \frac{(y-a_y)^2}{\sigma_y^2}
- \frac{2r(x-a_x)(y-a_y)} {\sigma_x \sigma_y}\right)\right] 
\end{equation}

The observational data from the Current BATSE GRB Catalog 
 will be used. There are 2702 GRBs, 
for {\bf 1956} of which both the  hardnesses and durations are measured.

In order to find the unknown constants in Eq.(\ref{gauss}) we use
the Maximum Likelihood ({\bf ML}) procedure of parameter estimation
\cite{bal03}. One can define the Likelihood
Function in the usual way, after fixing the value of $k$, in the
form
  $L=\sum \log p(x_i,y_i)$ ,
where $p(x_i,y_i)$ has the form given by Eq.(\ref{lfpr}).
Similarly, as it was done by Bal\'azs et al. \cite{bal03}, the EM
(Expectation and Maximization) algorithm is used to obtain the
$a_x, a_y, \sigma_x, \sigma_y, r$ and $p_l$ parameters at which
$L$ reaches its maximum value. We made the calculations for
different values of $k$ in order to see the improvement of $L$ as
we increase the number of parameters to be estimated.

The confidence interval of the parameters came from   
$ 2 (L_{max}-L_0) = \chi ^2_m $ equation.
  Moving from $k$=2 to 3 the number of parameters $m$ increases by 6 
(from 11 to 17), and $L_{max}$ grows from 1193 to 1237, which means a very 
low probability of being a chance ($10^{-10}$). 

Moving from $k$=3 to 4, the improvement in $L_{max}$ is only 6 (from 1237 to 1243), 
which can happen by chance with a probability of 6.2\%. Hence, the inclusion of 
the fourth class is not justified. Table 1. shows the parameters of the
best fit with $k$=3.


\begin{table}
\begin{tabular}{lrrrrrr}
\hline
\tablehead{1}{r}{b}{classes}
  & \tablehead{1}{r}{b}{$p_l$}
  & \tablehead{1}{r}{b}{$a_x=\log T_{90}$}
  & \tablehead{1}{r}{b}{$a_y=\log H_{32}$}
   & \tablehead{1}{r}{b}{\ \ \ \ \ \ \ \ $\sigma_x$ \ \ }
   & \tablehead{1}{r}{b}{\ \ \ \ \ \ \ \ $\sigma_y$ \ \ }
  & \tablehead{1}{r}{b}{r (corr.coef.)}   \\
\hline
 1   &    .245  &  -.301  &   .763  &   .525  &   .251  &   .163 \\
   2   &    .109  &   .637  &   .269  &   .474  &   .344  &  -.513 \\
   3   &    .646  &  1.565  &   .427  &   .416  &   .210  &  -.034 \\

\end{tabular}
\caption{Parameters of the Gaussian fits ($k=3$) in the $\{\log T_{90}; \log H_{32}\}$ 
plane. }
\label{tab:a}
\end{table}

\section{Results}

 The mathematical deconvolution of the 
$p(x,y)$ joint probability density of the observed quantities into Gaussian 
components does not necessarily mean that the physics behind the classes 
obtained mathematically is  different. It could well be possible that the 
true functional form of the distributions is not exactly Gaussian and that 
the algorithm of deconvolution  formally inserts a third one only in order 
to get a satisfactory fit.  One needs detailed investigations based on the 
physical (e.g. spectral) properties of the individual bursts to prove its 
astrophysical validity.

Norris et al. \cite{nor01} and Balázs et al. \cite{bal03} found 
compelling evidence that there is a significant 
difference between the short and long GRBs. This might indicate that different 
types of engines are at work. The relationship of long GRBs to the massive 
collapsing objects is now also observationally well established, 
and the relation between the comoving and observed 
time scales is well 
understood \cite{fox, rype}. The short bursts can be identified 
as originating from neutron star (or black hole) mergers. Therefore the 
mathematical classification of GRBs into the short and long classes is 
 physically justified.

An important question that must be answered in this context is whether the 
intermediate group of GRBs, obtained in the previous paragraph from the 
mathematical classification, really represents a third type of burst 
physically different from both the short and the long ones.

The classification into the short, intermediate and long classes is based 
mainly on the duration of the burst. From Table 1 one may infer that these 
three classes differ also in the hardnesses. The difference in the hardnesses 
between the short and long group is well known \cite{kou}. 
According to these data the {\bf intermediate} GRBs are the {\bf softest }
among the three 
classes. This different small mean hardness and also the different average 
duration suggest that the intermediate group should also be a different 
phenomenon, that is, both in hardness and in duration  the third group  
differs from the other two. On the other hand, no significant
correlation exists 
between the hardness and the duration within the short and the long classes. 
Thus, these two quantities may be taken as two independent variables, and 
the short and long groups are different in both these independent variables.

 In contrast, there is a strong {\bf anticorrelation} between the 
hardness and the 
duration within the intermediate class. This is a surprising, new result, and 
because the hardness and the duration are not independent in the third group, 
one  may simply say that only one significant physical quantity is responsible 
for  the hardness and the duration within the intermediate group. Consequently, 
the situation is quite different here, because one needs two independent 
variables to describe the remaining two other groups. This is a strong 
constraint in modeling the third group.  Hence, the question of the true 
nature of the  physics in the intermediate group remains open, and 
needs further analysis.

In this paper we have shown that statistically a third
group of GRBs exists. Also statistically no further groups are needed
to describe the $\{\log T_{90}; \log H_{32}\}$ 
distribution of bursts.
Finally, 11\%  of GRBs in the Current BATSE Catalog belong to 
the intermediate class.
The memberships of the cataloged bursts are available on the internet \cite{ho05}.

\begin{theacknowledgments}
  
Thanks are due to D. L. Band, C.-I. Bj\"ornsson, J. T. Bonnell, L.
Borgonovo, J. Hakkila, S. Larsson, P.
M\'esz\'aros, J. P. Norris, H. Spruit and G. Tusn\'ady  for
valuable discussions. This study was supported by OTKA, grant No.
 T048870. 

\end{theacknowledgments}


\bibliographystyle{aipproc}   

\bibliography{sample}

\IfFileExists{\jobname.bbl}{}
 {\typeout{}
  \typeout{******************************************}
  \typeout{** Please run "bibtex \jobname" to optain}
  \typeout{** the bibliography and then re-run LaTeX}
  \typeout{** twice to fix the references!}
  \typeout{******************************************}
  \typeout{}
 }

\end{document}